\documentclass[a4paper,fleqn,usenatbib]{mnras}
\usepackage{newtxtext,newtxmath}
\usepackage[T1]{fontenc}
\usepackage{ae,aecompl}
\usepackage{subfigure}
\usepackage{longtable, tabu}
\usepackage{color}
\usepackage{graphicx}	
\usepackage{amsmath}	
\usepackage{amssymb}	
\usepackage{color}
\title[Low energy electron impact of Ne II and Ne III ions]{Fine-structure electron-impact excitation of Ne$^+$ and Ne$^{2+}$ for low temperature astrophysical plasmas}

\author[Qianxia Wang et al.]{
 Qianxia Wang$^{1, 2}$
 S. D. Loch$^{1}$\thanks{lochstu@auburn.edu},
 Y. Li$^{1}$,
 M. S. Pindzola$^{1}$,
 R. S. Cumbee$^{3, 4}$,
\newauthor{P. C. Stancil$^{3}$,
 B. M. McLaughlin$^{5}$ and
C. P. Ballance$^{5}$\thanks{c.ballance@qub.ac.uk}}
\\
$^{1}$Department of Physics, Auburn University, Auburn, AL 36849, USA\\
$^{2}$Department of Physics and Astronomy, Rice University, Houston, TX 77005, USA\\
$^{3}$Department of Physics and Astronomy and Center for Simulational Physics, 
            University of Georgia, Athens, GA 30602, USA\\
$^{4}$NASA Goddard Space Flight Center, Greenbelt, MD 20771, USA\\
$^{5}$Centre for Theoretical Atomic and Molecular Physics (CTAMOP),
          School of Mathematics and Physics,\\
	 Queen's University Belfast, Belfast BT7 1NN, Northern Ireland, UK}

\date{Accepted March 2,2017. Received July 11, 2016; in original form}

\pubyear{2015}

\begin{document}
\label{firstpage}
\pagerange{\pageref{firstpage}--\pageref{lastpage}}
\maketitle

\begin{abstract}
Collision strengths for electron-impact of
the fine-structure levels within the ground term
of Ne$^+$ and Ne$^{2+}$ are calculated using the Breit-Pauli and
DARC $R$-matrix methods. Maxwellian-averaged effective collision strengths are
presented for each ion. The application of the current calculations is to very
low temperature astrophysical plasmas, down to 10 K, thus we examine the sensitivity
of the effective collision strengths to the resonance positions and underlying atomic structure. The use of the
various theoretical methods
allows us to place estimated uncertainties on the recommended effective
collision strengths. Good agreement is found
with previous $R$-matrix calculations at higher temperatures.
\end{abstract}




\section{Introduction}

Electron-impact fine-structure excitation of low charged ions is an important
cooling mechanism in most interstellar environments, especially in regions
with significant ionization fraction where electron-impact excitation is a 
strong populating mechanism for the excited states.
The lines from these fine-structure transitions can be observed from the infrared (IR)
to the submillimeter (submm) by a range of telescopes (e.g., the {\it Spitzer Space Telescope},
the Stratospheric Observatory for Infrared Astronomy (SOFIA), the {\it Herschel Space
Observatory}, the Atacama Large Millimeter Array (ALMA), etc.). Furthermore, fine-structure
excitation due to electron-impact is an important diagnostic tool for
the density, pressure, temperature, and/or ambient radiation field,
if sufficiently accurate rate coefficients can be obtained. Electron impact
fine-structure excitation has been studied fairly extensively for many ions
over the past few decades \citep{Pradhan1974,Butler1984,Johnson1987,Saraph1978,
Saraph1994,Griffin2001,Colgan2003,Berrington2005,Witthoeft2007,Burgos2009,
Ludlow2011,Malespin2011,McLaughlin2011,Wu2012,Auburn2013,Auburn2015} with $R$-matrix data being used
in most modeling databases.
However, almost all of these studies have primarily focused on high energies/high
temperatures relevant
to collisionally-ionized plasmas, thus much of the low temperature data being used has been extrapolsed from the available $R$-matrix data. Also, no detailed study has been taken on the uncertainty of the low temperature
rate coefficients. Thus, the aim of this paper is to produce a set of recommended fine structure rate coefficients for the ground term of Ne$^+$ and Ne$^{2+}$, along with an estimated set of uncertainties.

For plasmas of importance in this paper, we require rate coefficients
down to approximately 10~K, appreciating that achieving the accuracy 
in the underlying cross section at this temperature is difficult.   
Therefore it is important for astrophysical
models that collisional calculations are performed of sufficient accuracy at lower energies
for the low temperature rate coefficients. This will extend the available data down to lower temperatures
than exist currently in databases and will also improve the accuracy of the low temperature rate coefficients.

The fine-structure line emissions from Ne II and Ne III, populated via electron-impact excitation,
are observed in the IR and are known to be very important for probing
H II regions. Previous work \citep{Glassgold2007,Meijerink2008} proposed that Ne II and Ne III fine-structure
lines could serve as a diagnostic of the source of an
evaporative flow, as well as of signatures of X-ray irradiation,  the so-called X-ray dominated
regions (XDRs). This is because
hard X-rays have sufficient energy to generate multiple ionization states
of neon which can then be collisionally excited. 
$R$-matrix calculations have been available for these 
and neighbouring ion stages of Ne.
Specifically, collision strengths
for the 1s$^2$2s$^2$2p$^5$ ($^{2}$P$^0_{3/2}$) - ($^{2}$P$^0_{1/2}$)
transition of Ne$^+$ have been calculated using a $R$-matrix method via
the JAJOM approach \citep{Saraph1978,Johnson1987,Saraph1994}, that transforms 
$LS$-coupled K-matrices into level-level cross sections.
The collision strengths of the transitions among
levels of the lowest configurations for Ne$^{2+}$ were evaluated
by \cite{Pradhan1974} and \cite{Butler1984} using the IMPACT close-coupling code. 
McLaughlin et al. \citep{McLaughlin2000,McLaughlin2002,McLaughlin2011}
extended this approach to a large configuration-interaction 
representation of the target, supplemented by a few extra pseudo-orbitals
to improve the target description further.  

Here,
we have re-investigated these two Ne ions for several reasons. 
Previous work has focused primarily on higher electron-impact energies than considered here with only a few of
their Maxwellian averaged effective collision strengths going below 800 K. Thus, we investigate the
sensitivity of the very low temperate rate coefficients to changes in the atomic structure, threshold energies, and resonance positions.
This leads naturally to the second focus 
of the paper, which is the exploration of uncertainty in the rate coefficients at very low temperatures. 
To this end, two different theoretical level-resolved $R$-matrix approaches  have been applied: 
the Breit-Pauli (BP) approximation \citep{Berrington1987} and the fully relativistic
Dirac method \citep{Norrington1981,Dyall1989,Grant2007}.
Ostensibly, if the underlying electronic structure adopted in each 
approach was exactly the same there would be little expectation of differences
in the collision strengths. However, with the use of different atomic structure codes 
and the choices made in their use, this invariably leads to small differences in transition probabilities (Einstein
A coefficients) 
and subsequently, dynamical quantities such as collision strengths.  

Due to the low temperature focus of this paper, we are interested in the sensitivity of the effective collision strengths to
the threshold energy position, the target wave-functions, resonance positions, 
and anything that can affect the background cross section.
We appreciate that the height and position of a single resonance
can dramatically affect the results at these temperatures. 
We shall explore the variation in results to threshold energy and resonance positions
by calculating collision strengths where the target
energies have been shifted (or not) to NIST energies \citep{NIST2015}. Furthermore, we explore the sensitivity  
of the target wave-function via different target expansions within the BP $R$-matrix and DARC $R$-matrix methods. After investigating the differences
between all calculated effective collision strengths for the same transition,
a recommended dataset is determined for each ion.

\begin{figure}
\begin{center}
\includegraphics[height=85mm,width=80mm]{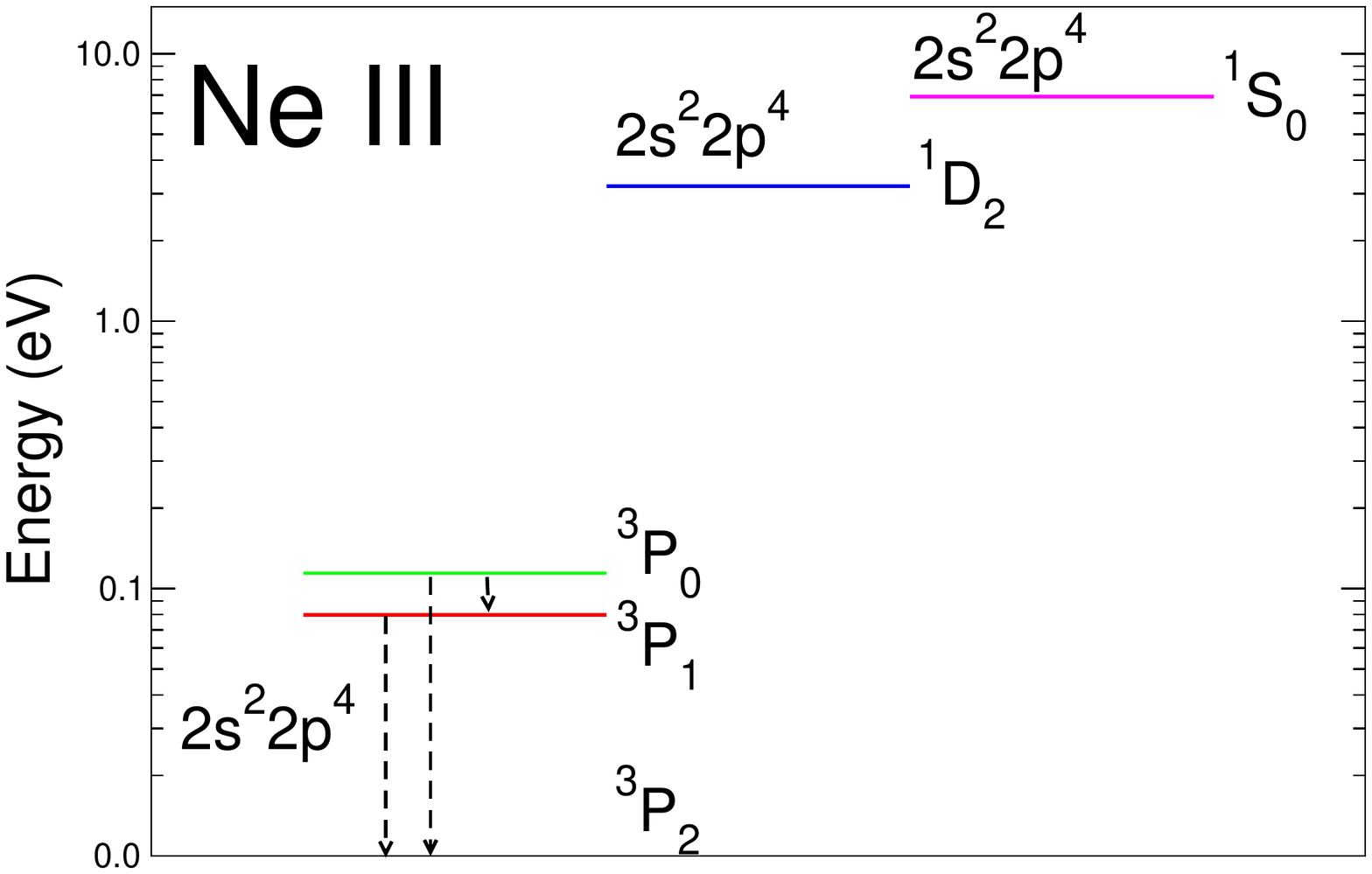}
   \caption{Energy diagram for Ne III. Three fine-structure transitions are shown:
   2s$^2$2p$^4$ ($^{3}$P$_1$) - ( $^{3}$P$_2$),
   2s$^2$2p$^4$ ($^{3}$P$_0$) - ($^{3}$P$_1$), and
   2s$^2$2p$^4$ ($^{3}$P$_0$) - ($^{3}$P$_2$).
   }
\label{Ne2+_dia}
\end{center}
\end{figure}

We focus on excitation at low temperatures (10 -2000 K) in this paper. So for Ne$^+$,
only rates for the transition between the two lowest levels 1s$^2$2s$^2$2p$^5$ ($^{2}$P$^0_{3/2}$) - ($^{2}$P$^0_{1/2}$)
are presented. Also, the transitions between the three lowest fine-structure levels of Ne$^{2+}$ (see energy diagram in Figure \ref{Ne2+_dia})
are investigated here.

%
%

The rest of this article is organized as follows. In Sec. 2,
we describe the theoretical methods used in this paper.
Sec. 3 presents the details of the calculations.
The calculated results, target energies, Einstein A coefficients, 
collision strengths and effective collision strengths 
for Ne$^+$ and Ne$^{2+}$ will be discussed in Sec. 4.
Sec. 5 provides a summary of the results.

\section {Theory}
Level-resolved electron-impact excitation cross-section calculations using $R$-matrix theory,
employs a similar formalism either in semi-relativistic (BP) or relativistic (DARC) implementations.
 Ne$^+$ and Ne$^{2+}$ are not highly charged, therefore both semi-relativistic
and fully relativistic methods are equally
applicable. The main differences arise from the choices made in the determination of the 
target orbitals for use in dynamical calculations. The atomic
structure code AUTOSTRUCTURE \citep{Badnell1986} generates non-relativistic orbitals whereas the
General Relativistic Atomic Structure Package (GRASP; \cite{Dyall1989,Grant2007}) formulates and diagonalises
a Dirac-Coulomb Hamiltonian to produce the relativistic orbitals. The former
is used in the BP/ICFT \citep{Berrington1995,Griffin1998}  R-matrix collisional calculations and the latter in the
Dirac Atomic $R$-matrix Code (DARC) \citep{Chang1975,Norrington1981,Grant2007} calculations to obtain
level-to-level electron impact excitation (EIE) cross sections.

The BP $R$-matrix method is a set of parallel codes developed 
from modified serial versions of the RMATRX~I codes \citep{Berrington1995}. 
Both the BP and ICFT models
recouple underlying $LS$ coupling calculations, the former transforms
several $LS$-resolved Hamiltonians into a $jK$-coupled Hamiltonian
as opposed to the ICFT approach that transforms unphysical $LS$-resolved 
$K$-matrices into level-resolved collision strengths. 
In general there has been very good agreement between the ICFT and BP
$R$-matrix methods \citep{Griffin1998,Burgos2009,McLaughlin2011}.

The implementation of various flavours of $R$-matrix theory
are used in this study. The review book of Burke \citep{Burke2011}
provides an excellent overview of non-relativistic ($LS$ coupling),
semi-relativistic (BP/ICFT) and relativistic (DARC) electron-impact excitation.
The comparison of BP and ICFT results benefits from the use 
of a completely consistent atomic structure calculation. Thus, the comparison with
the multiconfiguration Dirac-Fock (MCDF) structure results from GRASP and the resulting DARC
collision strengths, provides a means to investigate effects due to changes in the target structure.
In all cases every effort has been made to optimize the orbitals on the
fine-structure levels of the ground term. The DARC calculation employs relativistic 
orbitals from the initial atomic structure calculations throughout the remainder of the computation.
It should be restated that low temperature astrophysical constraints on both 
our 
Ne systems means we are pursuing only transitions between the fine-structure levels of the 
ground term, and that any excited states are included for the main purpose of improving the 
energy levels of those low-lying states through configuration interaction. Given that the energy 
separation between the ground state $n=2$ and the excited $n=3$ levels for either Ne$^+$ or Ne$^{2+}$
exceeds 2 Ryd., it is unlikely that Rydberg states attached to the $n=3$ levels would perturb our
$n=2$ results.

\section{calculation details}
\subsection{Target state calculation }
Given the low temperature focus of this paper, 
only small scale calculations are required for the fine-structure transitions
within the ground term.
Furthermore, we explore the variation of our results 
in relation to various configuration-interaction (CI) expansions.
Thus, we consider both a small and larger CI expansion for Ne$^+$ and
Ne$^{2+}$, with the configurations described in Table \ref{target}. The models are referred to 
as BP $n=2$, DARC $n=2$, BP $n=3$, and DARC $n=3$.

%
%
%

\begin{table*}
\begin{tabular}{|c | c | c | c | c |}
\hline 
Ne$^+$ DARC/BP $n=2$ & Ne$^+$ DARC/BP $n=3$   & Ne$^{2+}$ DARC/BP $n=2$ &Ne$^{2+}$ DARC/BP $n=3$\\ 
\hline                                                                                        
1s$^2$2s$^2$2p$^5$ & 1s$^2$2s$^2$2p$^5$      & 1s$^2$2s$^2$2p$^4$  & 1s$^2$2s$^2$2p$^4$  \\   
1s$^2$2s2p$^6$     & 1s$^2$2s2p$^6$          &  1s$^2$2p$^6$       &  1s$^2$2p$^6$       \\  
                   & 1s$^2$2s2p$^5$3$l$      &  1s$^2$2s2p$^5$     &  1s$^2$2s2p$^5$     \\ 
                   & 1s$^2$2s$^2$2p$^4$3$l$  &                     &  1s$^2$2s2p$^4$3$l$ \\ 
                   &                         &                     &  1s$^2$2s$^2$2p$^3$3$l$ \\ 
                   &                         &                     &  1s$^2$2p$^5$3$l$ \\				 
\hline                                                     
\end{tabular}    
\caption{Target expansions for Ne$^+$ and Ne$^{2+}$.} 
\label{target}       
\end{table*}

Orbitals are optimized automatically using the GRASP code \citep{Dyall1989,Grant2007}, which are then used 
in subsequent scattering calculation within the Dirac $R$-matrix method. We focus
on low energies, the orbitals of interest belong then to the first several terms. Scattering calculations for the DARC $n=3$ model
use orbitals obtained from GRASP from the DARC $n=2$ model, which are held fixed. This improves the energies and A-values of the levels in the ground term. We note if all orbitals are optimized simulataneously in a GRASP n=3 calculation this leads to much larger differences with NIST values. Thus, our GRASP n=2 and n=3 calculations for both ions are optimized for the fine-structure levels of interest to this work. In the GRASP calculations for each ion we used the option of extended average level (EAL) to optimize the energies.
 
For the BP scattering models, we optimize the target input orbitals  using AUTOSTRUCTURE \citep{Badnell1986} with different sets of parameters. We developed a code to vary the orbital scaling parameters used in AUTOSTRUCTURE, comparing the resulting energies and A-values with NIST values until a minimum was found in the differences (for the transitions of interest) with the NIST tabulated values \citep{NIST2015}. 
The optimized orbital scaling parameters are as follows; $\lambda_{1s}$=0.8, $\lambda_{2s}$=1.2,  and $\lambda_{2p}$=1.08 for BP $n=2$ 
calculations of Ne III.  For the BP $n=3$ calculations of Ne III:   $\lambda_{1s}$=0.8, $\lambda_{2s}$=1.2, 
 $\lambda_{2p}$=1.08, $\lambda_{3s}$=0.8, $\lambda_{3p}$=1.2, and $\lambda_{3d}$=0.8.  

Similarly, we use the same proceedure to generate parameters for the target orbitals for
Ne II  BP $n=2$ and $n=3$ collision models. However, in this case we find that the resulting effective collision strengths do not change much compared with the unoptimized structure results, we therefore adopted the unoptimized parameters for the target orbitals in our BP collision calculations for Ne$^+$ in this paper. It should be noted that these BP results for Ne II will not be used for our recommended data, and instead will be used only in estimating the uncertainty on our final data.

\subsection{Scattering calculation}
Details specific to the current $R$-matrix calculations are summarized  in Table \ref{parameter}, where we have  included
all the relevant parameters used in the different calculations. 
The radius of the $R$-matrix sphere for the different collision calculations is selected automatically during the $R$-matrix calculations.
The number of  continuum basis orbitals for each angular momentum and partial waves are chosen to converge the results for the low
temperature calculations. The energy mesh is selected to ensured resonances were fully resolved, particularly for the lowest 
temperatures and the  subsequent effective collision strengths. The energy ranges calculated for the collision strengths are determined by the
effective collision strengths temperature range of interest. It should also be noted that all of the energy levels in the target structure calculations were included in the scattering calculations.
%
%
%
\begin{table*}
\caption{Scattering calculation parameters used in our work on Ne II and Ne III ions for different target expansions.}
\label{parameter}         
\begin{tabular}{cccccc}
\hline  
                                        			&Ne II  $n=2$	& Ne II $n=3$  		& Ne III $n=2$ 		  &Ne III $n=3$\\ 
                                        			&  DARC, BP          	& DARC, BP             	& DARC, BP                &DARC, BP \\
\hline
Radius of $R$-matrix sphere (a.u.) 	&  5.40, 5.87      	& 19.83, 21.60         	& 4.89,5.24    		  &13.28, 14.35\\
Continuum basis orbitals for		&  20                 	& 20                    & 20                      &20             \\
each angular momentum    		&                      	&                       &                         &                  \\
Partial waves J                 			& 0 -- 20              	& 0 -- 20               & 0 -- 20                 &0 -- 20            \\
Energy mesh (Ryd.)  	          		& 2.5$\times$10$^{-6}$ 	& 2.5$\times$10$^{-6}$  & 3.125$\times$10$^{-6}$  &3.125$\times$10$^{-6}$ \\
Energy range (Ryd.)                    		& 0.007 -- 0.107     	& 0.007 -- 0.107        & 0.0058 -- 0.1658   	  &0.0058 --  0.1658 \\
Temperature range (K)  			& 10 -- 2000		& 10 -- 2000 		& 10 -- 2000		  &10 -- 2000\\ 
\hline                                                  
\end{tabular} 
\end{table*}

\subsection{Effective collision strength calculation}
The effective collision strength \citep{Seaton1953,Eissner1969} can be calculated from
the collision strengths via:
\begin{equation}
\Upsilon_{ij}=\int_0^\infty \Omega_{ij}\exp \biggl (\frac{-\epsilon_j }{kT_e} \biggr ) d \biggl (\frac{\epsilon_j}{kT_e} \biggr ),
\end{equation}
where $\Omega_{ij}$ is the collision strength for the transition
from level $i$ to $j$, $\epsilon_j$ is the energy
of the scattered electron, $T_e$ the electron temperature,
and $k$ Boltzmann's constant.

The Maxwellian excitation rate coefficient, $q_{ij}$, is used
widely in astrophysical modeling codes. The relationship
between $q_{ij}$ and $\Upsilon_{ij}$ is
\begin{equation}
q_{ij}=2\sqrt{\pi}\alpha ca^{2}_{0}\biggl (\frac{I_{\rm H}}{kT_e} \biggr )^{1/2}\frac{1}{\omega_i}e^{-\frac{\Delta E_{ij}}{kT_e}}\Upsilon_{ij},
\end{equation}
where $\alpha$ is the fine-structure constant, $c$ the speed of light, $a_0$ the Bohr radius, $I_{\rm H}$ the hydrogen ionization
potential, $\Delta E_{ij}$ the energy difference in the fine-structure levels, and $\omega_i$ the degeneracy in the lower level.
Compared with $q_{ij}$($T_e$), $\Upsilon_{ij} $($T_e$) is a smoother function and can be more accurately interpolated.                                                                                                                                                                       

\section{Results and discussion}
Astrophysical plasma modellers who study IR/submm observations of low
temperature plasmas, such as the interstellar medium, require atomic rate coefficients down to
temperatures as low as 10 K. This will place very stringent tests on the accuracy of the atomic structure 
and collisional calculations. The excitation rate coefficients will be very sensitive to small changes in
the atomic structure. As a result, the structure will impact the rate coefficients through changes in the
threshold energy, resonance strengths and positions, and changes in the background cross section.
For this reason, we have performed a range of 
calculations using different methods (BP $n=2$, DARC $n=2$, BP $n=3$, and DARC $n=3$). These will be
used to explore the variation of the effective collision strengths, particularly at
low temperatures. The purpose of including the $n=3$ configurations is to improve the energies
and transition probabilities for the levels within the ground term.

\subsection {Bound-state energies and radiative rates for Ne$^+$ and Ne$^{2+}$}
Our recommended dataset shall be the model that minimizes the difference between
the calculations and NIST A-values and level energies \citep{NIST2015}.
The results are shown in Tables \ref{energy} and \ref{A-value}.
The percent error ($\delta\%$) shown is
calculated by $\frac{x-x_{NIST}}{x_{NIST}}$$\times$$100\%$
with the NIST data providing the accepted values.

%
%
%
%
\begin{table*}
\caption{Fine-structure energy levels for Ne II and Ne III (in Ry) compared to the NIST values \citep{NIST2015}. The configurations and terms listed in the first two columns label different levels.
	Column 3  gives the corresponding energies from the tabulated NIST values.  The percent error after each theoretical energy indicates the deviation from the NIST value.
	The last line for each ion in the  table is the average error $\delta\%$ of each theoretical calculation.} 
\label{energy}      
\begin{tabular}{c  c  c  c  c c c  c  c  c c}
\hline 
Configuration   &  Term ($^{2s+1}L_J$)	& NIST 	& BP $n=2$ 	&$\delta\%$	&  BP $n=3$ &$\delta\%$ &DARC $n=2$	&$\delta\%$	& DARC $n=3$ &$\delta\%$\\ 
\hline                                                                                                                                                                                                                                                                
Ne II 2s$^2$2p$^5$ 	&  $^{2}$P$^0_{3/2}$	&0.0000	&  0.0000  	&   0   		&  0.0000  	&  0     		& 0.0000 	& 0     		&  0.0000 	&0    \\                             
             	  	&  $^{2}$P$^0_{1/2}$	&0.0071	&  0.0069  	&  2.82 	&  0.0070  	&  1.91  		& 0.0076 	& 7.2   	&  0.0079 	&11.3 \\   
\hline            
Avg. $\delta\%$     &                 			&      		&          	&  1.41 	&          	& 0.96   		&        		&  3.6 	&          	&5.65  \\
\hline
Ne III 2s$^2$2p$^4$  &  $^{3}$P$_2$  	& 0.0000 	& 0.0000	& 0       	&  0.0000	&  0   	 		& 0.0000   	&  0     	&  0.0000   	& 0    \\   
                  		&  $^{3}$P$_1$  		& 0.0059 	& 0.0057    &  3.39       	&  0.0057	&  3.39 		& 0.0060   	&  1.59  	&  0.0061   	& 3.39 \\  
                  		&  $^{3}$P$_0$  		& 0.0084 	& 0.0084    &  0       	&  0.0083	&  1.19 		& 0.0088   	&  4.33  	&  0.0090   	& 7.14 \\ 
                  		&  $^{1}$D$_2$  		& 0.2355 	& 0.2513   &  6.70     	&  0.2557	&  8.58 		& 0.2521   	&  7.05  	&  0.2470   	& 4.88\\ 
                  		&  $^{1}$S$_0$  		& 0.5081 	& 0.4942    & 2.76       	&  0.4914	&  3.29 		& 0.4795   	&  5.62  	&  0.4734   	& 6.83\\ 
\hline                                                     
Avg. $\delta\%$   &               			&       		&        		&  2.57       	&        		& 3.29  		&          	&  3.72  	&            	& 4.45 \\
\hline 
\end{tabular}
\end{table*}

%
%
%
\begin{table*}
\caption{Einstein A coefficients (in s$^{-1}$)  for magnetic-dipole-transitions (M1),  within the same configuration, for the Ne II and Ne III ions compared 
               to the NIST values \citep{NIST2015}. Columns are as  in Table 3.}  
\label{A-value}     
\begin{tabular}{c  c  c  c  c c c  c  c  c  c }
\hline 
Ion& Transition&    NIST &   BP $n=2$  &    $\delta\%$  &  BP $n=3$  			&  $\delta\%$  	&  DARC $n=2$ &  $\delta\%$ &  DARC $n=3$ & $\delta\%$\\ 
\hline                                                                                                                         
Ne II    		&2s$^2$2p$^5$ ($^{2}$P$^0_{3/2}$) - ($^{2}$P$^0_{1/2}$) 	& 8.59E-3 		& 7.84E-3 	& 8.68  & 8.07E-3 & 6.1  & 8.16E-3 & 5.01 & 8.14E-3& 5.24 \\
\hline
Ne III	 		&2s$^2$2p$^4$ ($^{3}$P$_2$) - ($^{3}$P$_1$)             		& 5.84E-3 		& 5.47E-3     & 6.34  & 5.42E-3 & 7.19 & 5.86E-3 & 0.42 & 5.81E-3& 0.51\\
       			&2s$^2$2p$^4$ ($^{3}$P$_1$) - ($^{3}$P$_0$)             		& 1.10E-3 		& 1.38E-3     & 25.45  & 1.36E-3 & 23.64& 1.15E-3 & 4.95 & 1.14E-3& 3.64\\
\hline                                                     
Avg. $\delta\%$   &               									&       			&        		  &15.90 &             &15.42  &              & 2.69 &            	& 2.08 \\
\hline 
\end{tabular}     
\end{table*}       

AUTOSTRUCTURE \citep{Badnell1986} and GRASP \citep{Dyall1989,Grant2007} are two different atomic strucuture codes used to generated target energies and orbitals
for the BP and DARC R-matrix methods repectively. They give rise to 3 and 108 levels for Ne$^+$, 
and 10 and 226 levels for Ne$^{2+}$ for $n=2$ and $n=3$ target expansions.
The energies for the levels within the ground term are presented in Table \ref{energy} and the associated A-values in Table \ref{A-value}.
In general, the percentage errors show that the agreement between theoretical and NIST values is reasonable.
 For Ne$^+$, the average percentage error for the BP $n=2$, BP $n=3$, DARC $n=2$, and
DARC $n=3$ target expansions are 1.41\%, 0.96\%, 3.6\% and 5.65\%, respectively. 
For Ne$^{2+}$, the corresponding average percent errors for target energies are 2.57\%, 3.29\%, 3.72\% and 4.45\%.

  
Table \ref{A-value} presents the comparison of Einstein A coefficients of the magnetic-diple (M1) for both the Ne$^+$ and Ne$^{2+}$ transitions. Due to the reasonable agreement for all the calculations with NIST energies, we use the A-values as our main selection criteria in recommending a final dataset for Ne$^+$ and Ne$^{2+.}$.
In both cases, the GRASP code produces closer agreement with NIST A values, compared with AUTOSTRUCTURE. 
The accuracy of the Einstein A coefficient depends on both the precision
of the target energies and the reliability of the target wave-functions.
The A values produced by GRASP n=2 and n=3 are close, because they use the same orbitals.
Overall the n=2 GRASP calculation has slightly better agreement with NIST A values and energies than the GRASP n=3 calculation, thus our recommended dataset for both ions is the DARC $n=2$ calculation. The other calculations can be used to gauge the variation between the different calculations, and will be used to produce an uncertainty estimate on our recommended data.

\subsection{Collision strengths and effective collision strengths for Ne$^+$ and Ne$^{2+}$ }
To our knowledge, there are no experimental results for the collision strengths 
for transitions within the ground complex for either of these ion stages.
Our goal is to determine the variation in effective collision strengths
between our best models as we progress to the very low temperatures required 
by the astrophysical applications. As stated above, the DIRAC n=2 effective collision strengths will be our recommended data, with the some of the other calculations being used to provide an uncertainty estimate. We have considered two different  
approaches to calculating meaningful representative uncertainties in our work.

In the first approach we calculate a percentage uncertainty on
the effective collision strengths simply using the standard deviation
of our three most accurate models as determined from the accuracy of
the energy levels and the associated A-values, given by
\begin{equation}
\%\Delta=\frac{{\sigma(\bar{x}_{best})}}{x_i}\times100\%
\label{eqn_stdev}
\end{equation}
where ${\sigma(\bar{x}_{best})}$ is the standard deviation. 
Secondly, we obtain a percentage difference comparing results
from our semi-relativistic and fully-relativistic $R$-matrix methods employing exactly 
the same set of non-relativistic target configurations. In this case, the
percentage difference is calculated by the formula $\frac{x_1-x_2}{(x_1+x_2)/2}$$\times$100$\%$.

\begin{figure}
\begin{center}$
\begin{array}{c}
\includegraphics[width=80 mm]{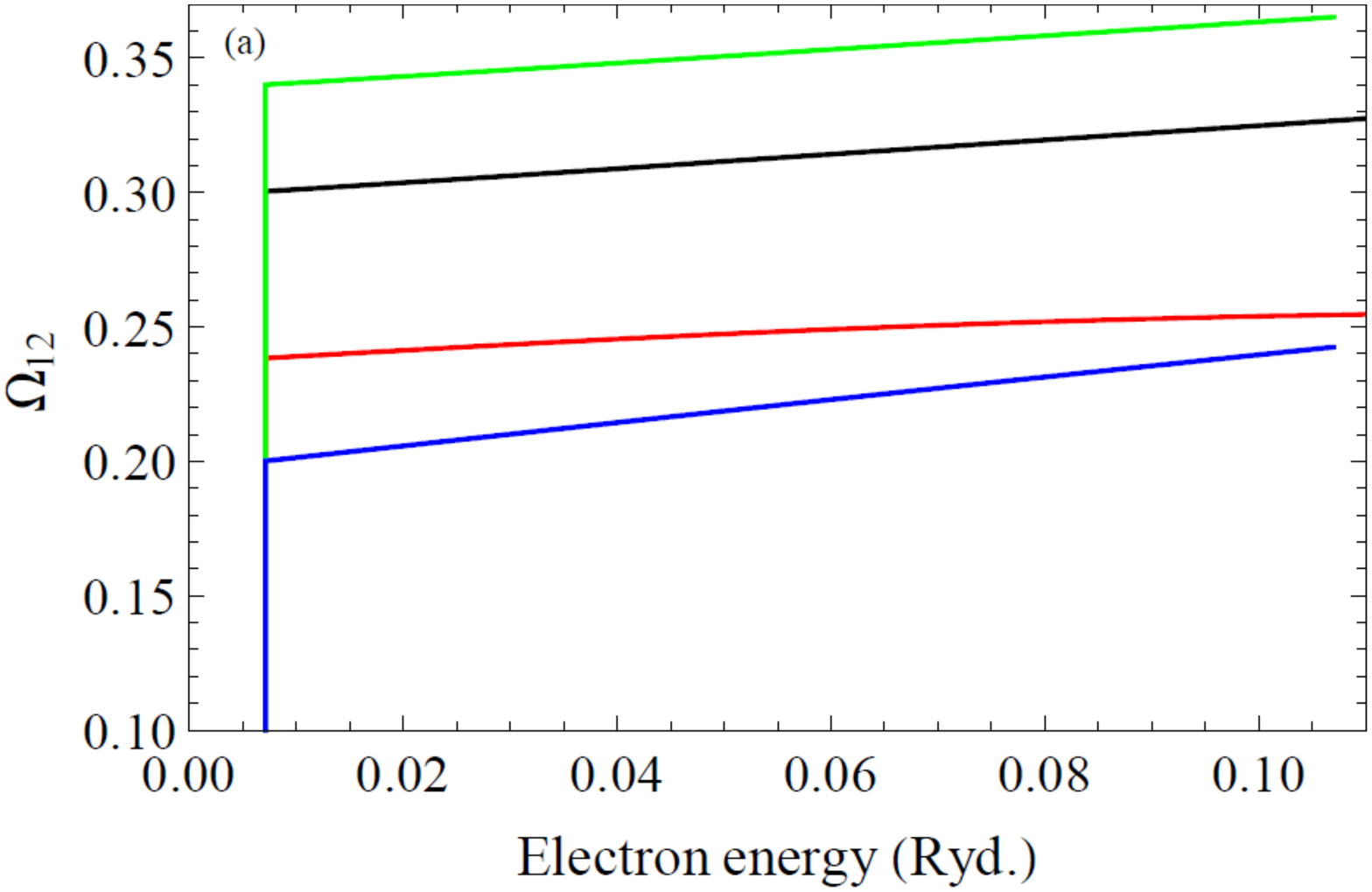} \\
\includegraphics[width=80 mm]{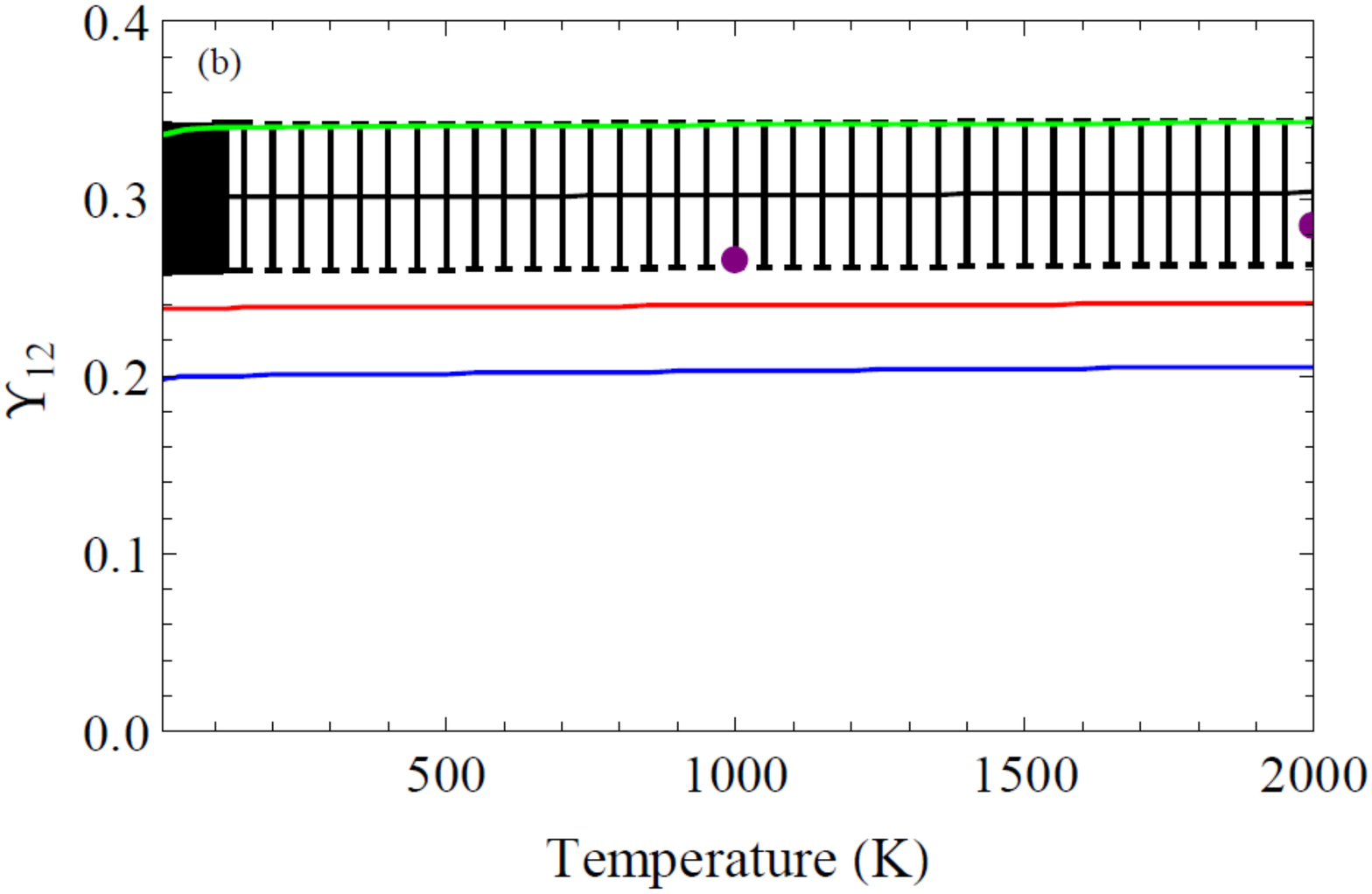}
\end{array}$
\end{center}
   \caption{Comparison of Ne II collision strengths (a) and effective collision strengths (b)
   for the 2s$^2$2p$^5$ $(^{2}$P$^0_{3/2}$) - ($^{2}$P$^0_{1/2})$
   transition between different target expansions: DARC $n=2$ (black line),
   DARC $n=3$ (red line), BP $n=2$ (green line), and BP $n=3$ (blue line). Uncertainty
   estimates are given for our recommended DARC $n=2$ results with comparison to the
   previous $R$-matrix calculation (purple circles) \citep{Griffin2001}.
   }
\label{Ne+}
\end{figure}

\begin{figure}
\begin{center}$
\begin{array}{c}
\includegraphics[width=80 mm]{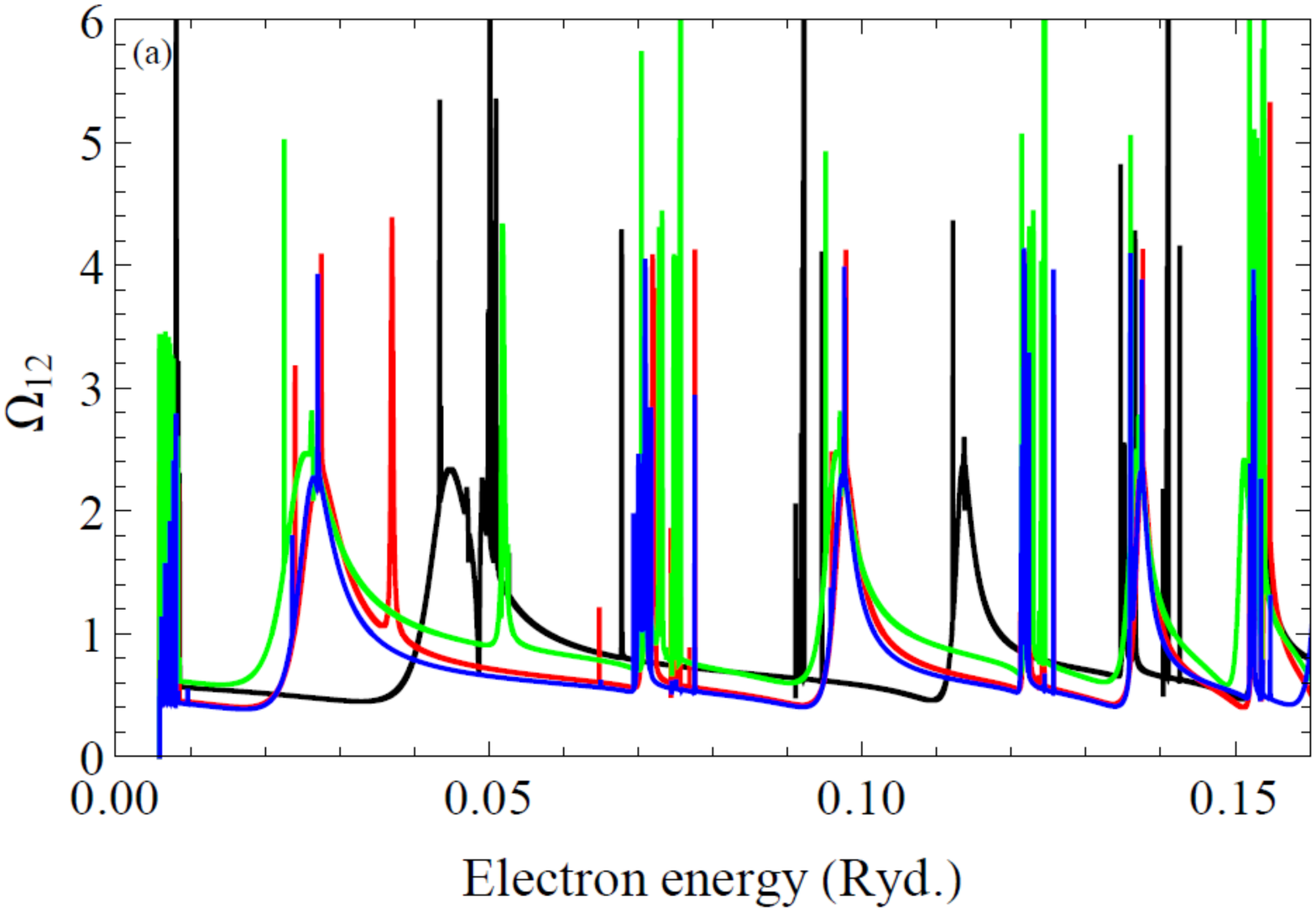} \\
\includegraphics[width=80 mm]{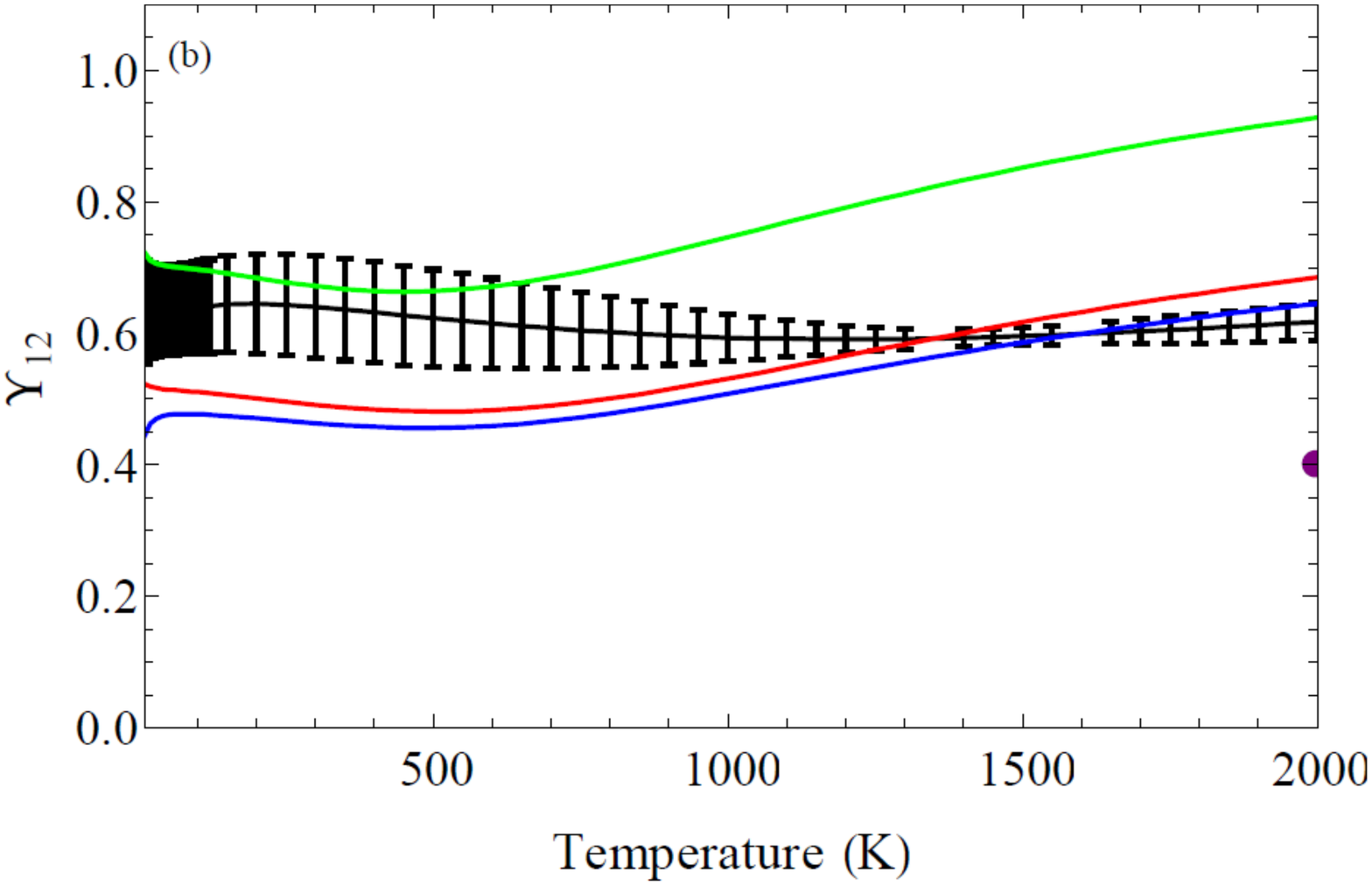}
\end{array}$
\end{center}
   \caption{Comparison of Ne III collision strengths (a) and effective collision strengths (b)
   for the 2s$^2$2p$^4$ ($^{3}$P$_2$) - ($^{3}$P$_1$)
   transition between different target expansions:  DARC $n=2$ (black line), DARC $n=3$
   (red line), BP $n=2$ (green line) and BP $n=3$ (blue line).  Uncertainty
   estimates are given for our recommended DARC $n=2$ results with comparison to the
   previous $R$-matrix calculation (purple circle) \citep{McLaughlin2011}.
   }
\label{Ne2+_1-2}
\end{figure}

\begin{figure}
\begin{center}$
\begin{array}{c}
\includegraphics[width=80 mm]{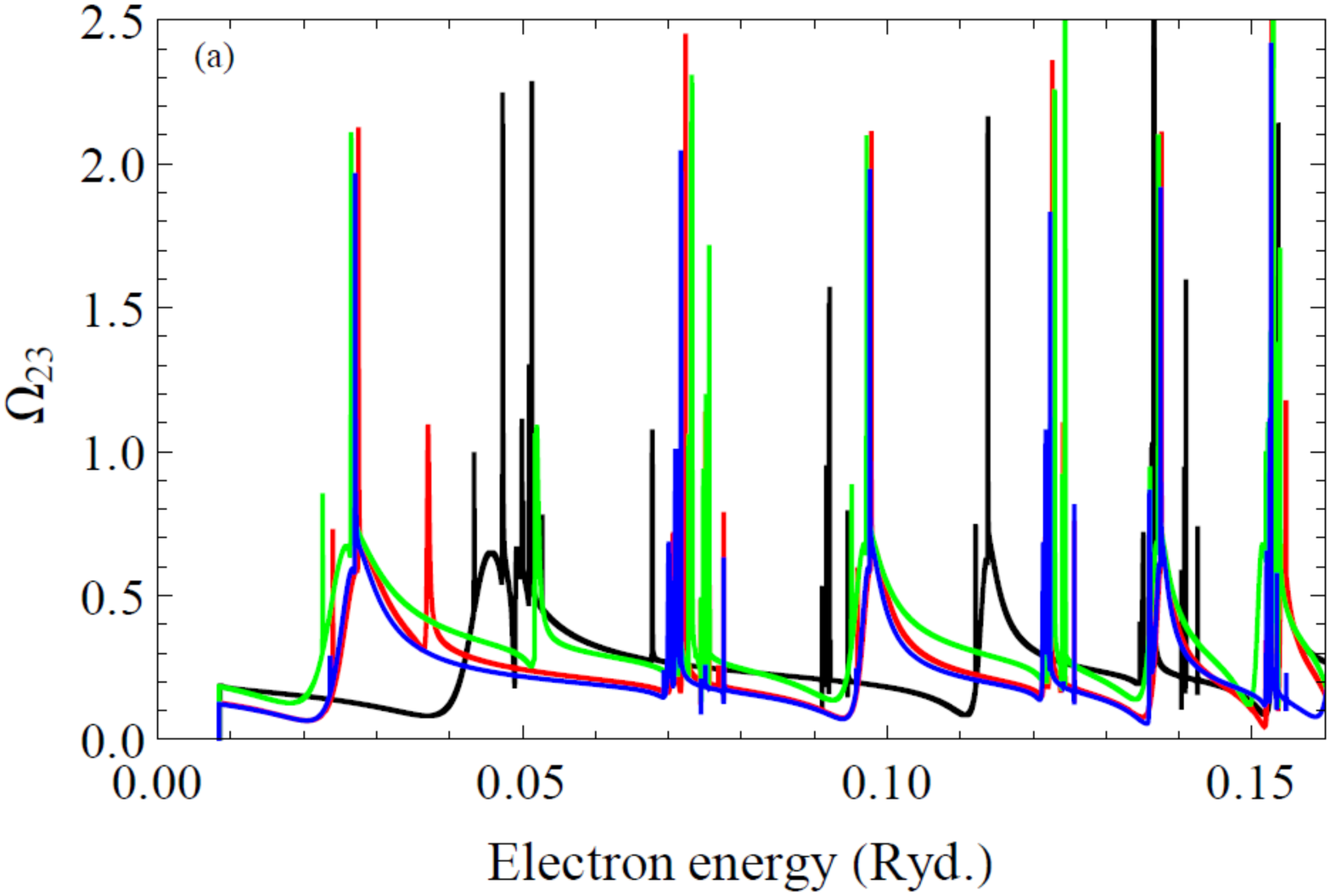} \\
\includegraphics[width=80 mm]{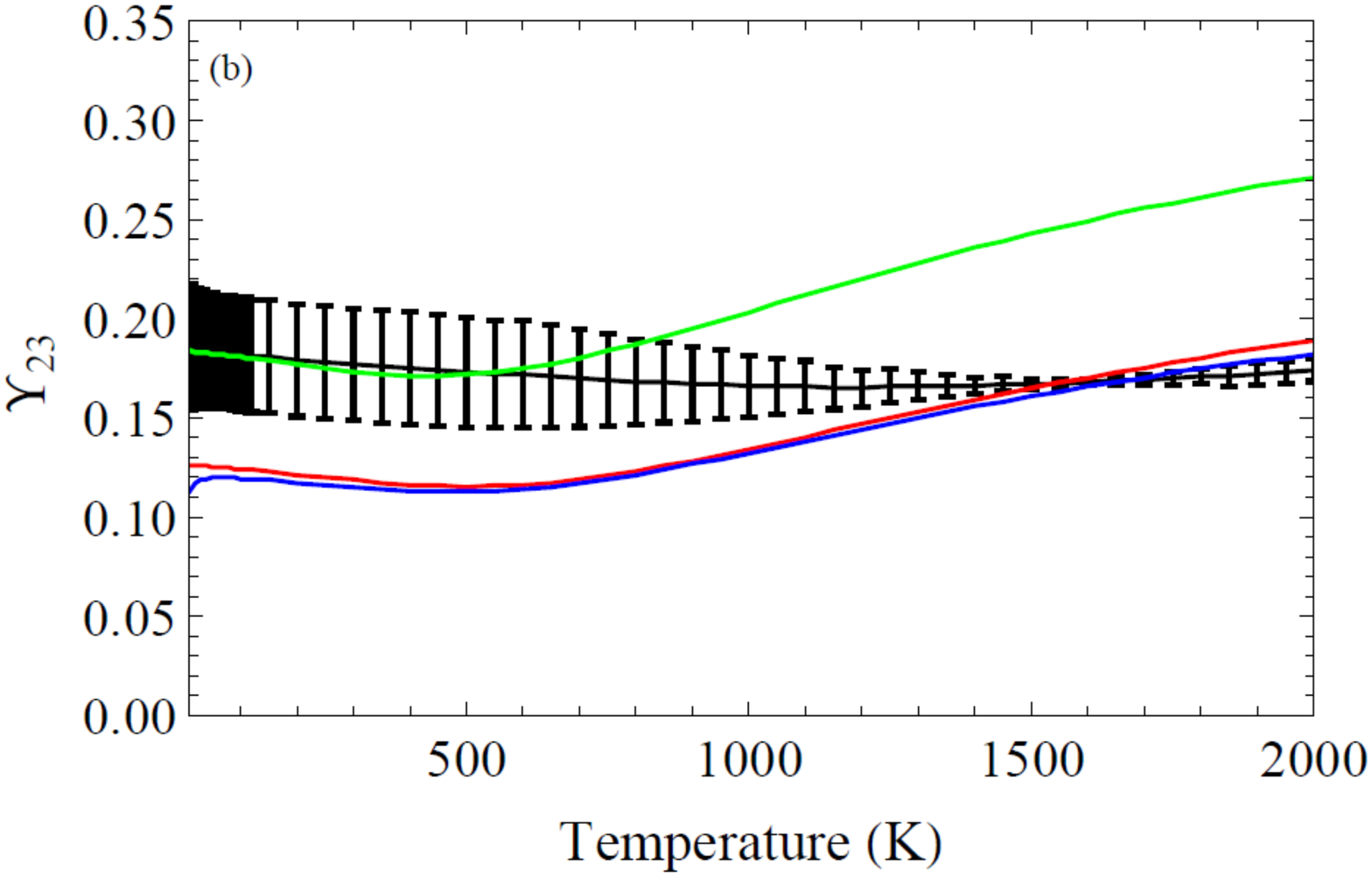}
\end{array}$
\end{center}
   \caption{Comparison of Ne III collision strengths (a) and effective collision strengths (b)
   for the 2s$^2$2p$^4$ ($^{3}$P$_1$) - ($^{3}$P$_0$)
   transition between different target expansions: DARC $n=2$ (black line), DARC $n=3$
   (red line), BP $n=2$ (green line) and BP $n=3$ (blue line).  Uncertainty
   estimates are given for our recommended DARC $n=2$ results.
   }
\label{Ne2+_2-3}
\end{figure}

\begin{figure}
\begin{center}$
\begin{array}{c}
\includegraphics[width=80 mm]{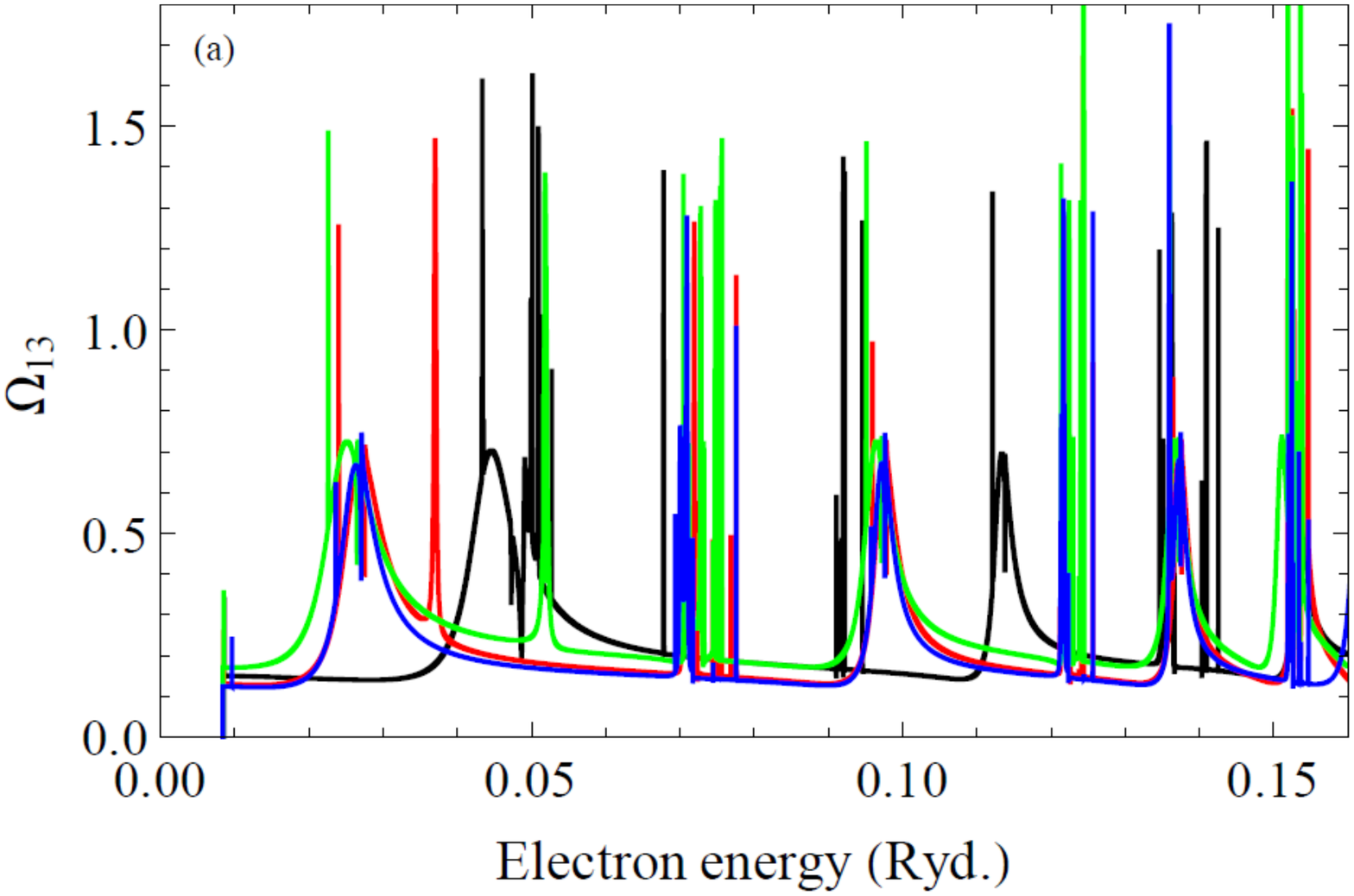} \\
\includegraphics[width=80 mm]{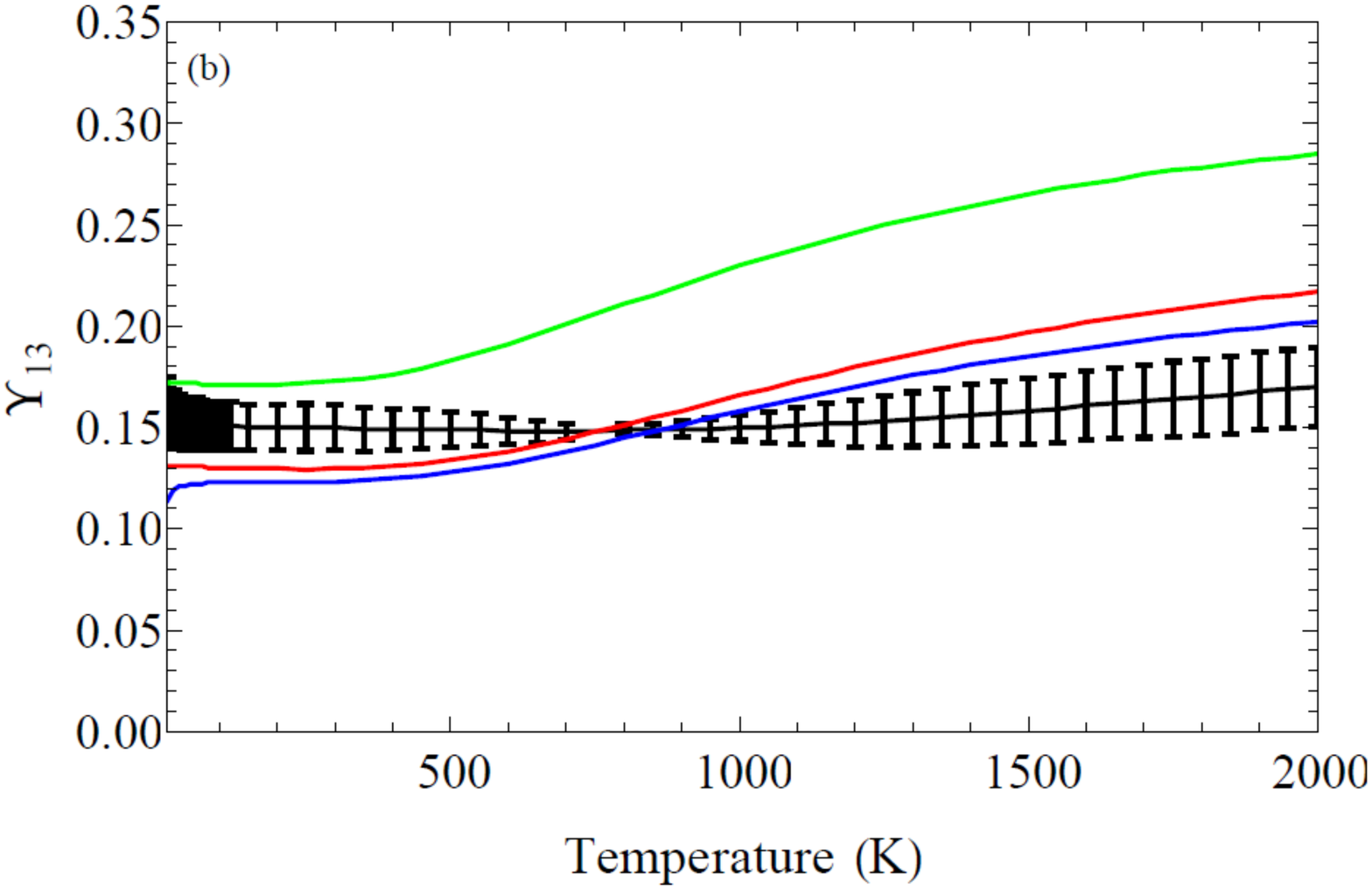}
\end{array}$
\end{center}
   \caption{Comparison of Ne III collision strengths (a) effective collision strengths (b)
   for the 2s$^2$2p$^4$ ($^{3}$P$_2$) - ($^{3}$P$_0$)
   transition between different target expansions: DARC $n=2$ (black line), DARC $n=3$
   (red line), BP $n=2$ (green line) and BP $n=3$ (blue line).
Uncertainty
   estimates are given for our recommended DARC $n=2$ results.
   }
\label{Ne2+_1-3}
\end{figure}

\begin{figure}
\begin{center}$
\begin{array}{c}
\includegraphics[width=80 mm]{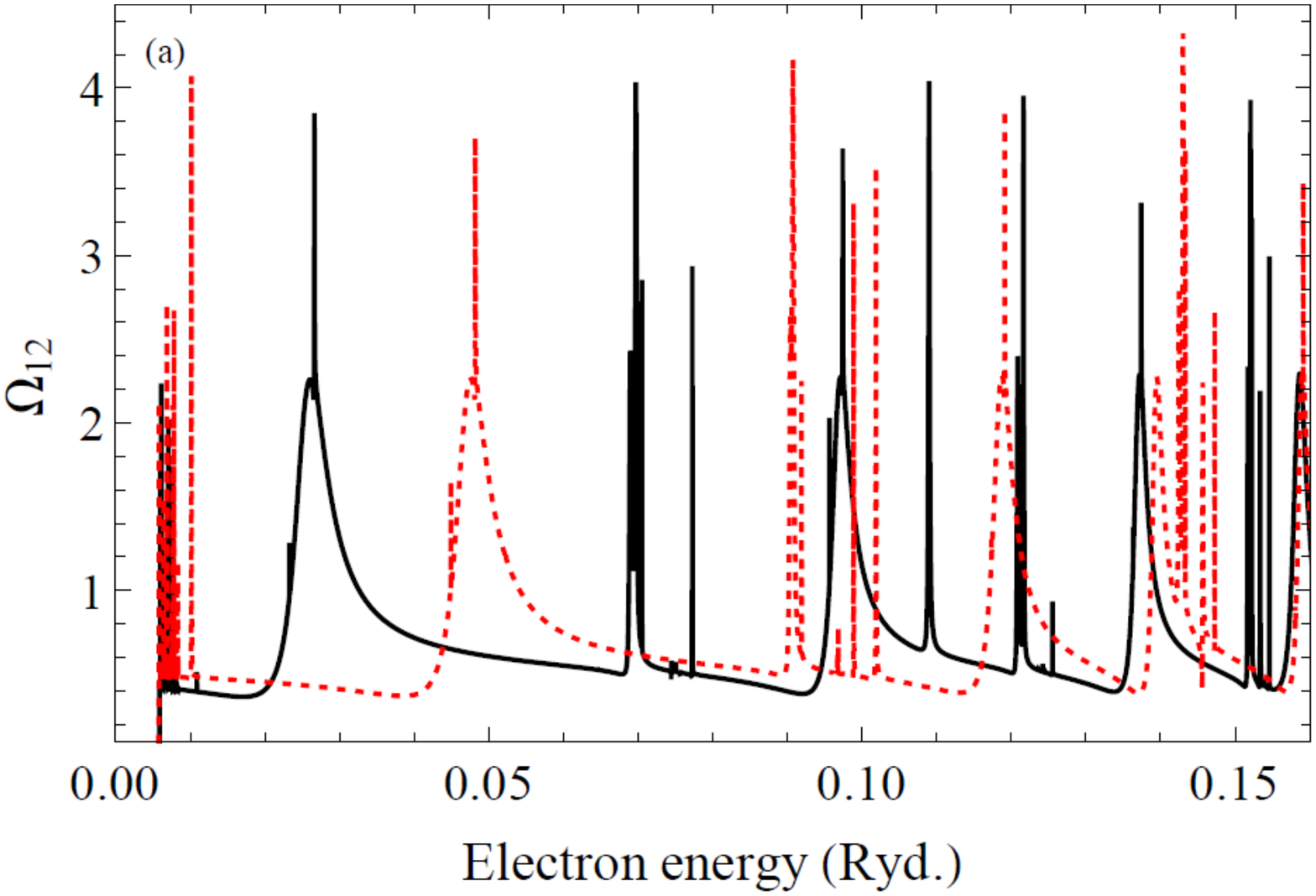} \\
\includegraphics[width=80 mm]{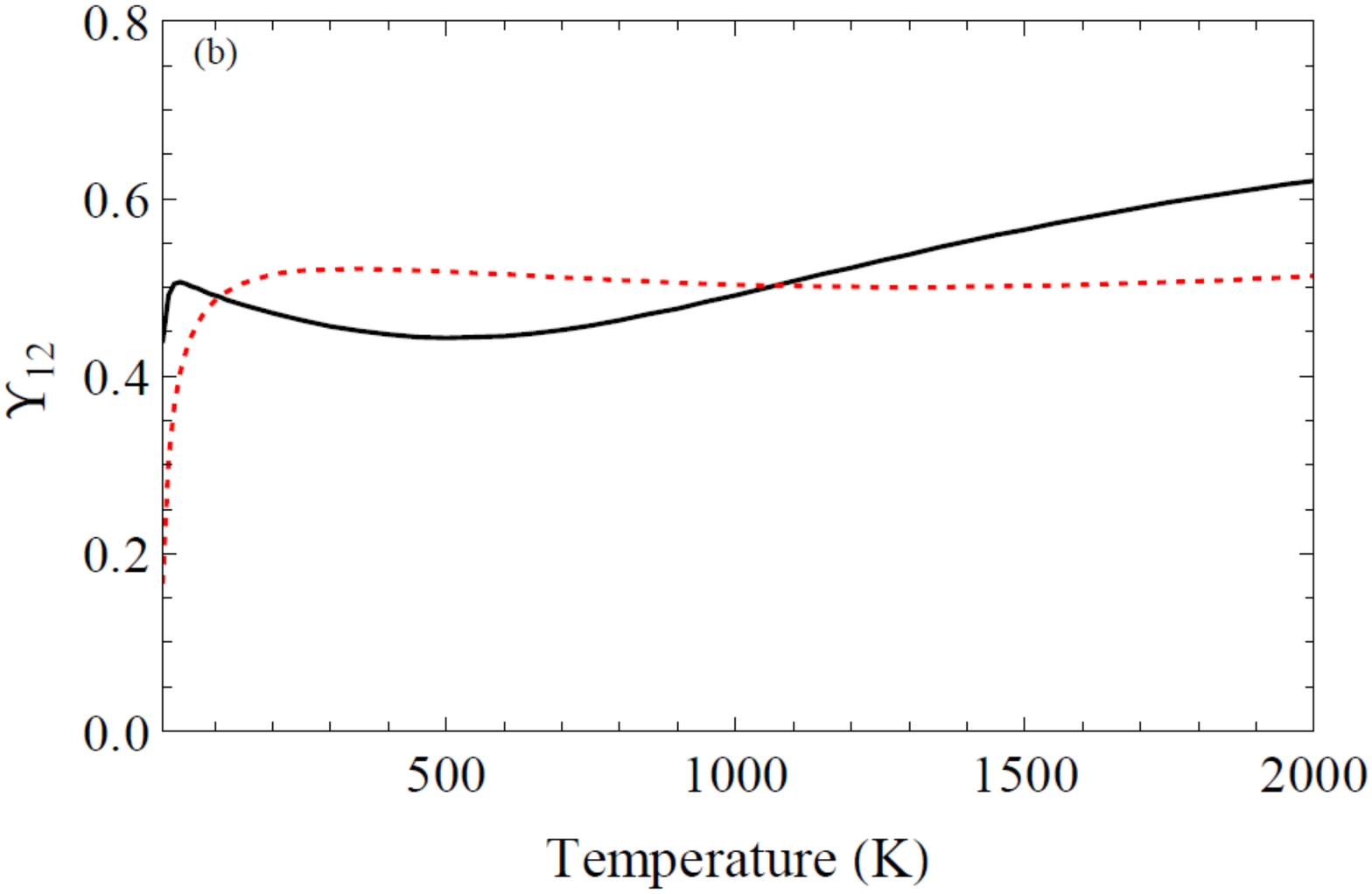}
\end{array}$
\end{center}
   \caption{Comparison of Ne III collision strengths (a) and effective collision strengths (b)
   for the 2s$^2$2p$^4$ ($^{3}$P$_2$) - ($^{3}$P$_1$)
   transition: BP $n=3$ with (black solid line) and without (red dotted line) the energy shift.
   }
\label{shift_Ne2+_1-2}
\end{figure}

\begin{figure}
\begin{center}$
\begin{array}{c}
\includegraphics[width=80 mm]{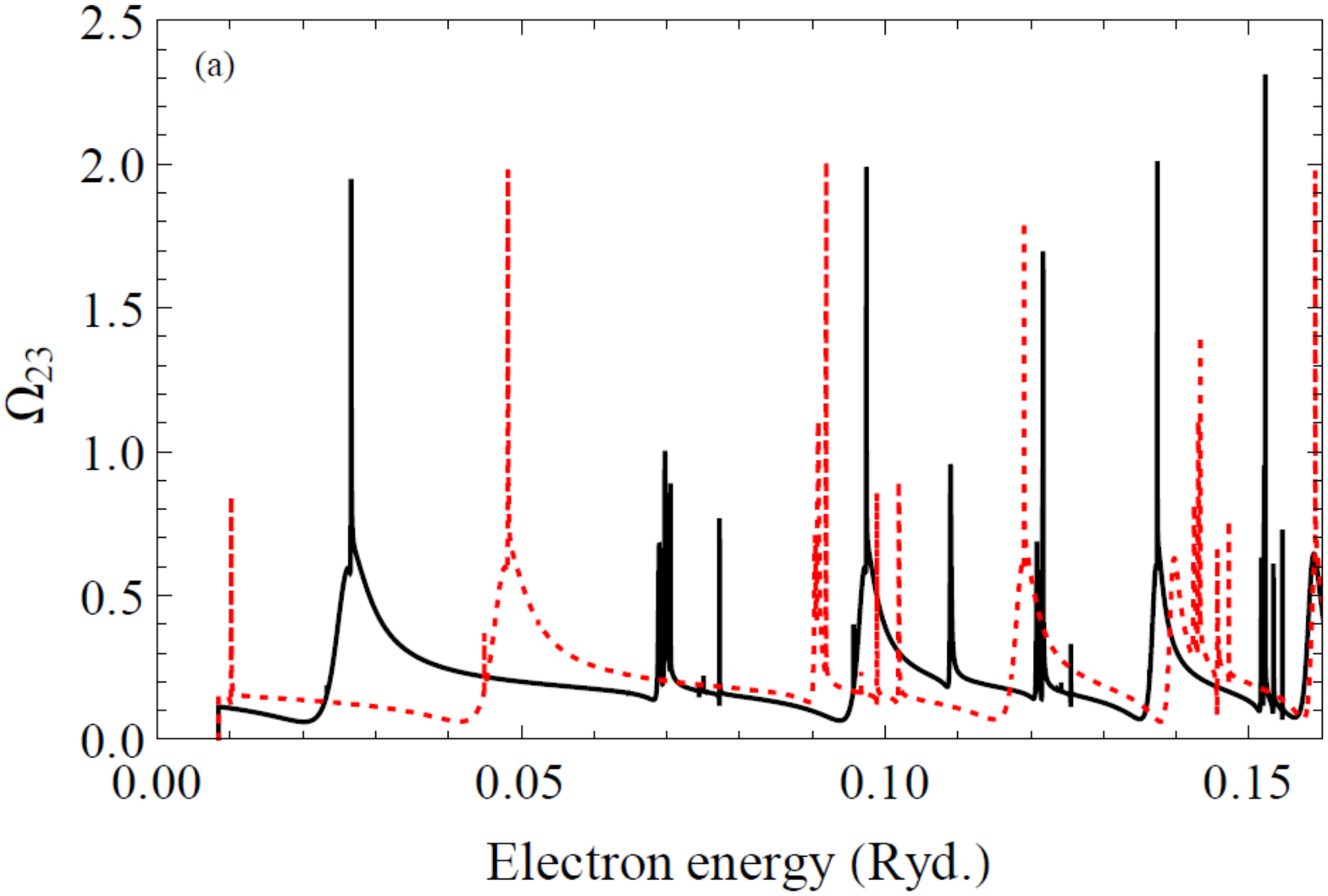} \\
\includegraphics[width=80 mm]{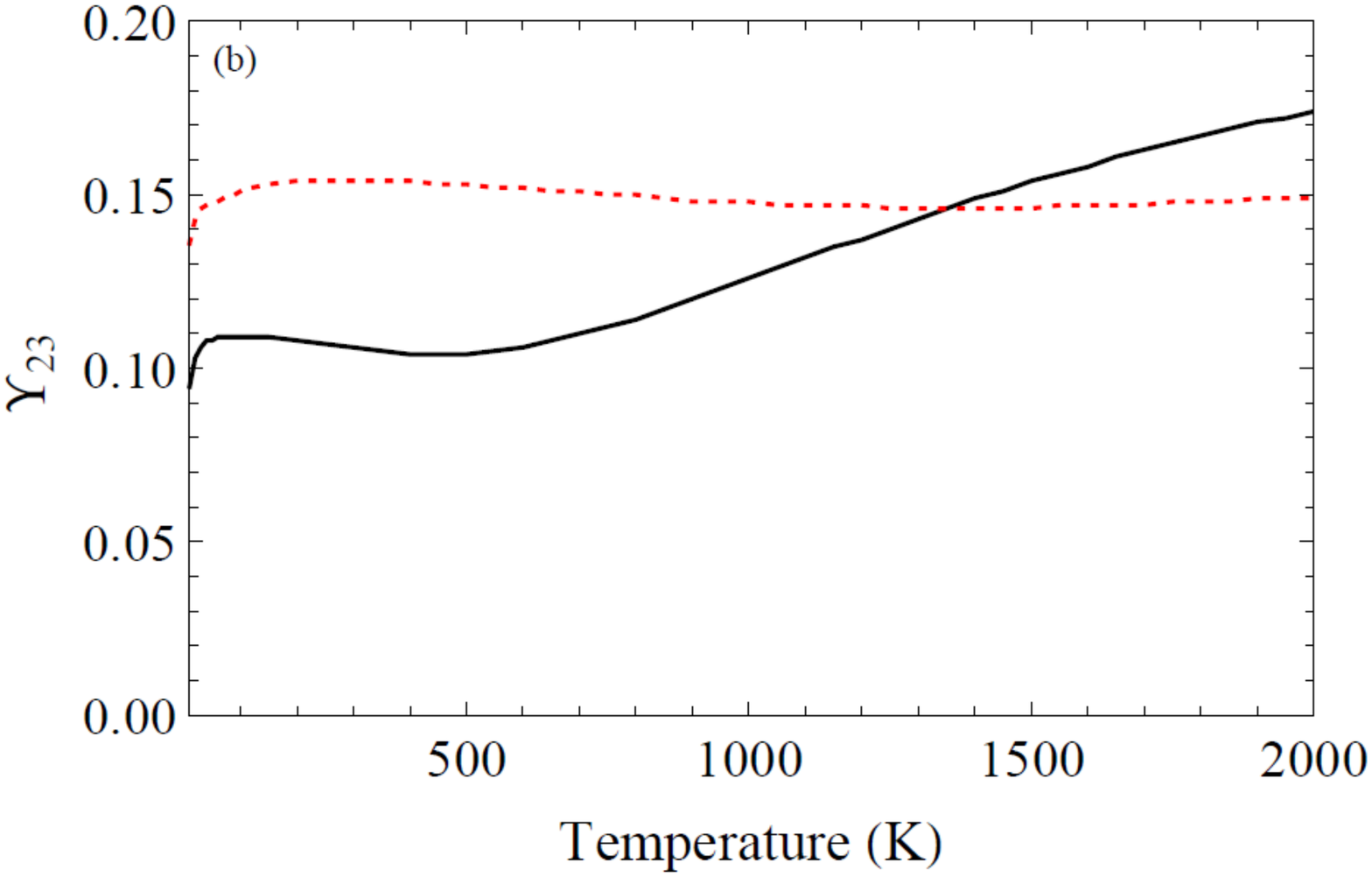}
\end{array}$
\end{center}
   \caption{Comparison of the Ne III collision strengths (a) and effective collision strengths (b)
   for the 2s$^2$2p$^4$ ($^{3}$P$_1$) - ($^{3}$P$_0$)
   transition: BP $n=3$ with (black solid line) and without (red dotted line) the energy shift.
   }
\label{shift_Ne2+_2-3}
\end{figure}

\begin{figure}
\begin{center}$
\begin{array}{c}
\includegraphics[width=80 mm]{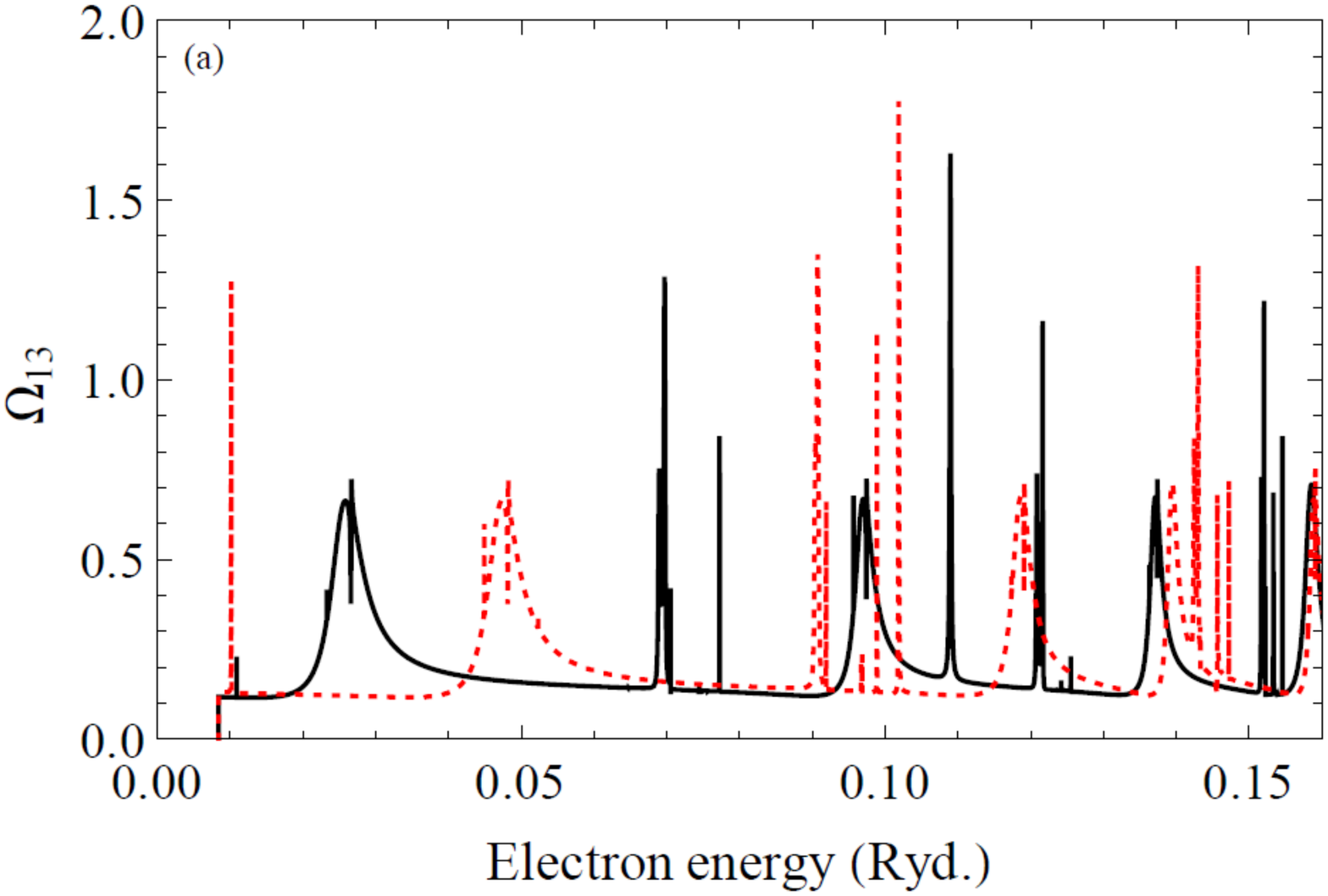} \\
\includegraphics[width=80 mm]{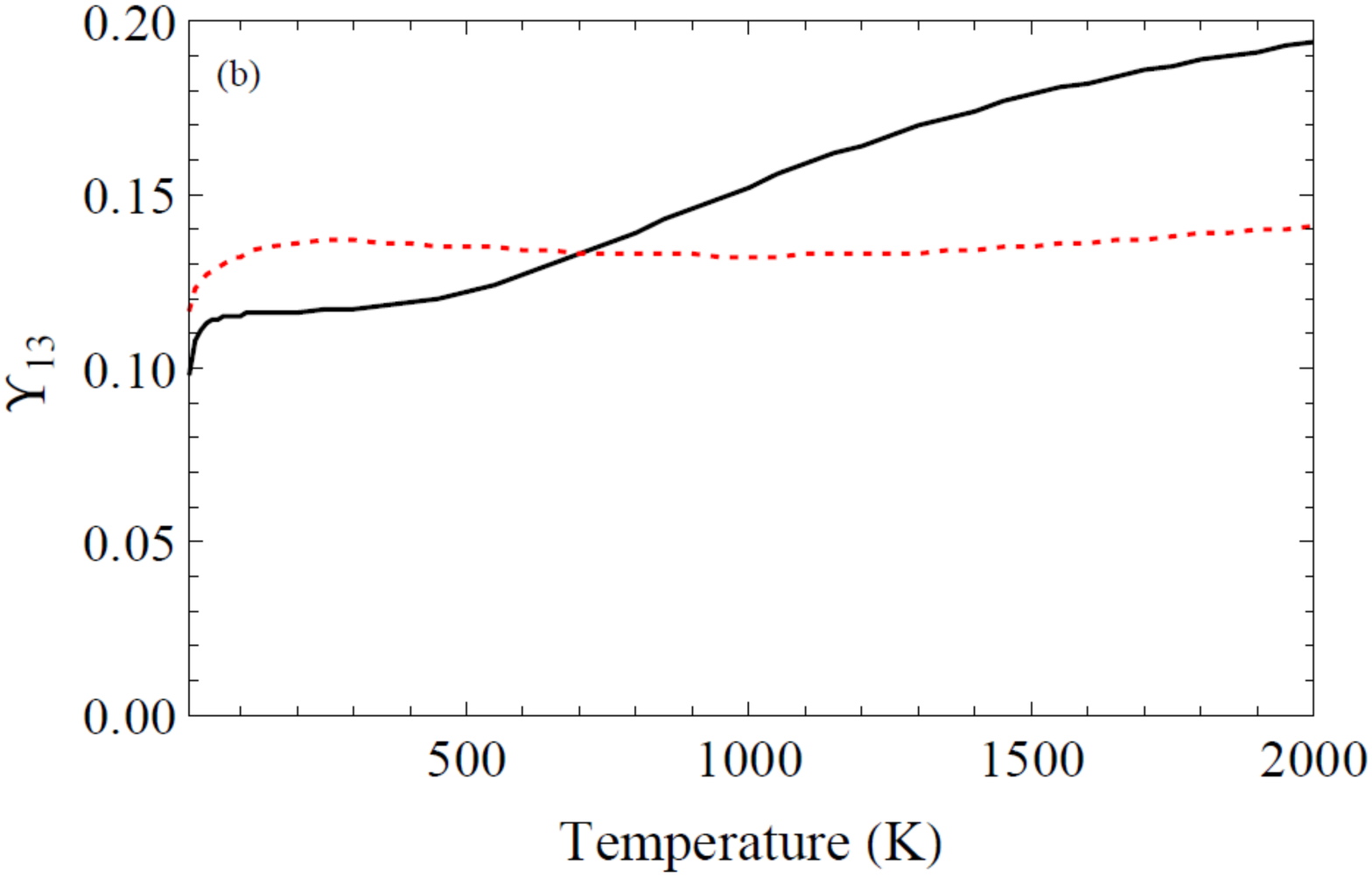}
\end{array}$
\end{center}
   \caption{Comparison of Ne III collision strengths (a) and effective collision strengths (b)
   for the 2s$^2$2p$^4$ ($^{3}$P$_2$) - ($^{3}$P$_0$)
   transition:  BP $n=3$ with (black solid line) and without (red dotted line) the energy shift.
   }
\label{shift_Ne2+_1-3}
\end{figure}

Figures \ref{Ne+} -- \ref{Ne2+_1-3} illustrate the collision strengths and effective collision
strengths for the fine-structure transitions of both Ne$^+$ and Ne$^{2+}$,
using the different $R$-matrix methods. In the evaluation of these effective collision strengths we use only our best calculations (i.e. optimized $\lambda$s where possible and with threshold energies shifted to NIST values). It is, however, interesting to investigate the effect of shifting to NIST energies on the effective collision strengths, thus
figures \ref{shift_Ne2+_1-2} -- \ref{shift_Ne2+_1-3} explore the effects of shifting
the target threshold energies to NIST values using the BP n=3 calculation for Ne$^{2+}$.

%
%

%
%

%
%

%
%

Sampling a range of calculations allows us to more objectively explore
the variation of collision strength in regards to the size of the
different CI expansions. As stated earlier, the sizeable energy 
separation between the $n=2$ and $n=3$ levels precludes the 
possibility of interloping resonances attached to the $n=3$ 
levels perturbing the cross sections from transitions amongst the 
$n=2$ levels. The influence of resonance contributions to 
effective collision strengths is only expected for the case of Ne$^{2+}$
due to the 2p$^4$ subshell supporting 3 levels within the ground term, whereas 
the resonances attached to the upper $J=\frac{1}{2}$ levels of Ne$^{+}$(2p$^5$)
lie in the elastic cross section of the Ne$^{+}$ ground state.  

Figure \ref{Ne+} shows collision strengths (top) and effective collision
strengths (bottom) for the Ne$^+$ 2s$^2$2p$^5$($^{2}$P$^0_{3/2}$) - ($^{2}$P$^0_{1/2}$)
transition. The largest collision strengths are from the BP $n=2$ calculation, the
next lower one is from the
DARC $n=2$ calculation, then from DARC $n=3$ and  BP $n=3$.
The DARC $n=2$ calculation is our recommended data set based upon 
A-value comparisons with the NIST database values \citep{NIST2015}. Furthermore, 
in the absence of experiment, the MCDF approach 
would be our recommended theoretical model. Subsequent effective
collision strengths were generated from the respective collision strengths
of each calculation. We note that beyond the current work,
a previous large-scale BP $R$-matrix calculation for Ne$^+$ 
has been carried out by \cite{Griffin2001}. However, the focus of that work was to provide 
a large comprehensive data set across a wide range of temperatures, but not
at the very low temperatures required by our study.
At 1000 and 2000 K, the DARC $n=2$ effective collision strengths are 
in reasonable agreement with this previous work. Our effective 
collision strengths are  0.302 at 1000 K and 0.304 at 2000 K,
compared with 0.266 and 0.286 from \citep{Griffin2001}, as shown in Fig. \ref{Ne+}, giving differences
of 12.7\% and 6.1\%, respectively. Thus, this supports our 
independent conclusion that our DARC $n=2$ effective collision strengths
should be the recommended dataset at even lower temperatures.
The recommended effective collision strengths are given 
in Table \ref{database}.  These results can also be obtained in
the formats of the Leiden Atomic and Molecular Database (LAMDA, \cite{Schoier2005}) and Stout \cite{stout}.

%
%
%
\begin{table*}
\caption{Effective collision strengths $\Upsilon_{12}$ and the uncertainty \%$\Delta$ for Ne II and Ne III ions
		calculated by the DARC approach using the $n=2$ target expansion.}
\label{database}         
\begin{tabular}{c  c  c  c  c c  c  c  c }
\hline 
Temperature&Ne II 2s$^2$2p$^5$($^{2}$P$^0_{3/2} $) -($^{2}$P$^0_{1/2}$)
&Ne III 2s$^2$2p$^4$($^{3}$P$_2$) - ($^{3}$P$_1$)
&Ne III 2s$^2$2p$^4$($^{3}$P$_2$) - ($^{3}$P$_0$)
&Ne III 2s$^2$2p$^4$($^{3}$P$_1$) - ($^{3}$P$_0$)\\ 
  (K)   &$\Upsilon_{12}$, \%$\Delta$ &  $\Upsilon_{12}$, \%$\Delta$ & $\Upsilon_{13}$, \%$\Delta$    &$\Upsilon_{23}$, \%$\Delta$ \\    
\hline                                                                                                                                                                                                                                                                
  10  	& 0.300, 13.99 	& 0.633, 12.24 	& 0.157, 11.26 	& 0.186, 17.09 \\                             
  40  	& 0.300, 13.74 	& 0.635, 10.76 	& 0.152, 8.50  	& 0.184, 15.83 \\
  70  	& 0.300, 13.74 	& 0.636, 10.70 	& 0.151, 8.03  	& 0.183, 15.62 \\
  100  	& 0.300, 13.74 	& 0.639, 10.92 	& 0.151, 7.88  	& 0.182, 15.71 \\
  150  	& 0.301, 13.82 	& 0.644, 11.45 	& 0.150, 7.63  	& 0.181, 15.65 \\
  300  	& 0.301, 13.69 	& 0.640, 12.14 	& 0.150, 7.63  	& 0.177, 16.01 \\
  450  	& 0.301, 13.69 	& 0.627, 12.00 	& 0.149, 6.54  	& 0.174, 16.14 \\   
  600  	& 0.301, 13.57 	& 0.615, 11.15 	& 0.148, 4.46  	& 0.172, 15.63 \\   
  750  	& 0.302, 13.67 	& 0.604, 9.53  	& 0.148, 2.23  	& 0.169, 13.68 \\
  900  	& 0.302, 13.52 	& 0.597, 7.55  	& 0.149, 2.59  	& 0.167, 11.15 \\
 1050  & 0.302, 13.52 	& 0.592, 5.38  	& 0.150, 5.19  	& 0.166, 8.53  \\
 1200  & 0.302, 13.52 	& 0.591, 3.52  	& 0.152, 7.62  	& 0.165, 5.62  \\ 
 1350  & 0.302, 13.40 	& 0.592, 2.23  	& 0.155, 9.14  	& 0.166, 3.35  \\ 
 1500  & 0.303, 13.50 	& 0.596, 2.17  	& 0.158, 10.32 	& 0.167, 1.49  \\
 1650  & 0.303, 13.36 	& 0.601, 2.88  	& 0.162, 10.84 	& 0.169, 1.28  \\
 1800  & 0.303, 13.36 	& 0.607, 3.64  	& 0.165, 11.40 	& 0.171, 2.15  \\
 1950  & 0.303, 13.36 	& 0.615, 4.28  	& 0.169, 11.39 	& 0.173, 3.30  \\ 
 2000  & 0.304, 13.46 	& 0.617, 4.52  	& 0.170, 11.53 	& 0.174, 3.52  \\ 
\hline                                                                                                                                                         
\end{tabular} 
\end{table*}

Employing the average percentage uncertainty given by Equation~(\ref{eqn_stdev}),
the $\bar{x}_{best}$ values used to calculate the uncertainty are the BP $n=3$, the DARC
$n=2$ and $n=3$ results, providing an uncertainty from 13 -- 14$\%$ for our recommended
DARC $n=2$ effective collision strengths, as shown in Figure \ref{Ne+}.

It is also of interest to consider the differences between the DARC and BP
calculations, for the cases when they both have the same configurations.
The differences of the effective collision strengths
between the DARC $n=2$ and BP $n=2$ are 11 -- 13$\%$, while the DARC $n=3$ and
BP $n=3$ differ by 16  -- 18 $\%$.

Considering next Ne$^{2+}$, Figures \ref{Ne2+_1-2} -- \ref{Ne2+_1-3} present the collision strengths (top) and effective collision strengths
(bottom) for three different transitions within the ground term, namely,
the ($^{3}$P$_2$) - ($^{3}$P$_1$) (Fig. \ref{Ne2+_1-2}),
($^{3}$P$_1$) - ($^{3}$P$_0$) (Fig. \ref{Ne2+_2-3}), and
($^{3}$P$_2$) - ($^{3}$P$_0$ (Fig. \ref{Ne2+_1-3}) transitions.
The collision strengths of different calculations have similar backgound for each of these transitions.
However, the resonance positions are shifted for each calculation, which cause the observed difference in the effective collision strengths.
 Previous calculations by \cite{McLaughlin2011},
which extended down to 2000 K, appears to be consistent with our recommended data, the DARC $n=2$ result.
The difference between our DARC $n=2$ and \cite{McLaughlin2011} are attributed
to the fact that that our  DARC $n=2$ calculation was focused on generating accurate
data only for fine structure transitions within the ground term, while \cite{McLaughlin2011}
results were focused on higher temperatures and higher n-shells in addition to the levels
within the ground term. Again, our recommended collision strength is that produced by the DARC $n=2$ 
calculation, based upon energy level, A-value comparisons with NIST data \citep{NIST2015} and published work,
as discussed above.

The uncertainty in the DARC $n=2$ results are again provided in a similar fashion using 
Eq. \ref{eqn_stdev} and the standard deviation of the other BP and DARC models. Values for $\bar{x}_{best}$ for Ne$^{2+}$ are
taken from the DARC $n=2$, $n=3$  and BP $n=3$ calculations.
Considering the effective collision strengths involving the higher excited state 
transitions (Figs. \ref{Ne2+_2-3} and \ref{Ne2+_1-3}), the DARC $n=2$ model remains our recommended dataset, with 
uncertainties given by the previously applied method. The uncertainty of the effective collision
strengths from the DARC $n=2$ calculations are 2 -- 12$\%$ (Fig. \ref{Ne2+_1-2}), 1-- 17$\%$ (Fig. \ref{Ne2+_2-3}), and
2 -- 12$\%$ (Fig. \ref{Ne2+_1-3}).

To investigate the sensitivity of the results due to shifting our target energies to NIST values, we consider in Figs. \ref{shift_Ne2+_1-2} -- \ref{shift_Ne2+_1-3} the effect of these shifts on the collision strengths 
for the BP n = 3 Ne III calculation.
The difference between the two BP calculations (with/without energy shift) is up to 89$\%$ for
the ($^{3}$P$_2$) - ($^{3}$P$_1$) transition, up to 39$\%$ for    
the ($^{3}$P$_1$) - ($^{3}$P$_0$), and up to 32$\%$ for 
the ($^{3}$P$_2$) -($ ^{3}$P$_0$). 
The large difference in the first transition is due to the presence of near threshold resonances.  
Thus, it is clearly important to include such shifts to NIST in the calculation of accurate low temperature rate coefficients. For this reason all of the data in our recommended dataset, and the data used for the uncertainty estimates, includes such NIST shifts.  In general, any system that has near threshold resonances would be particularly sensitive to such shifts, and this should be considered in future calculations of low temperature fine structure rate coefficients.

%
%

%
%

%
%

\section{Summary}
We calculated collision strengths and effective collision strengths
for Ne$^+$ and Ne$^{2+}$
with BP and DARC $R$-matrix methods. We are interested in the rates 
at low temperature (10 -- 2000 K),
so we focus on small energies (0.007 -- 0.107 Ryd. for Ne$^+$
and 0.0058 -- 0.1658 Ryd. for Ne$^{2+}$) and perform small scale
$R$-matrix calculations. After comparing the energies, the Einstein A coefficient (A$_{ij}$),
collision strengths ($\Omega_{ij}$) and effective collision strengths ($\Upsilon_{ij}$),  we conclude
that the DARC $n=2$ model provides the most reliable collision strengths
and effective collision strengths with the Einstein A coefficients generated
by this method being closest to the recommended values (i.e., NIST \cite{NIST2015}). Further, effective collision strengths computed with the DARC $n=2$ approach \citep{Griffin2001,McLaughlin2011} result in rates which agree well with
the existing data at higher temperatures calculated by large-scale $R$-matrix methods.

\section{Acknowledgement}

This work was funded under NASA grant NASA-NNX15AE47G. BMMcL  would like to thank Professors S. D. Loch, M. S. Pindzola and  Auburn University for their hospitality during recent research visits.





\bibliographystyle{mnras}                       
\bibliography{ne-ions}


\bsp	
\label{lastpage}
\end{document}